\documentclass[twocolumn,3p,times]{elsarticle}

\usepackage{ecrc}

\volume{00}

\firstpage{1}

\journalname{Nuclear Physics B Proceedings Supplement}

\runauth{Y. Takeuchi}

\jid{nuphbp}

\jnltitlelogo{Nuclear Physics B Proceedings Supplement}

\usepackage{amssymb}

\usepackage[figuresright]{rotating}

\usepackage{graphicx}

\begin{document}

\begin{frontmatter}


\title{Results from Super-Kamiokande}
\author{Y. Takeuchi for the Super-Kamiokande collaboration}
\ead{takeuchi@phys.sci.kobe-u.ac.jp}
\address{Dept. of Physics, Graduate School of Science, Kobe University, 1-1 Rokkodai-cho, Nada, Kobe 657-8501}

\dochead{}





\begin{abstract}

The recent results from Super-Kamiokande (SK) are reported.
On atmospheric neutrino analysis, we have performed a full 3-flavor
oscillation analysis with SK-I+II+III data. 
A CPT violation study on atmospheric neutrino is also done with 
SK-I+II+III data. 
On solar neutrino analysis, a 3-flavor oscillation analysis with
SK-III data is performed.

\end{abstract}


\end{frontmatter}


\section{Super-Kamiokande detector}

\begin{figure}
 \begin{center}
  \includegraphics[width=7cm,clip]{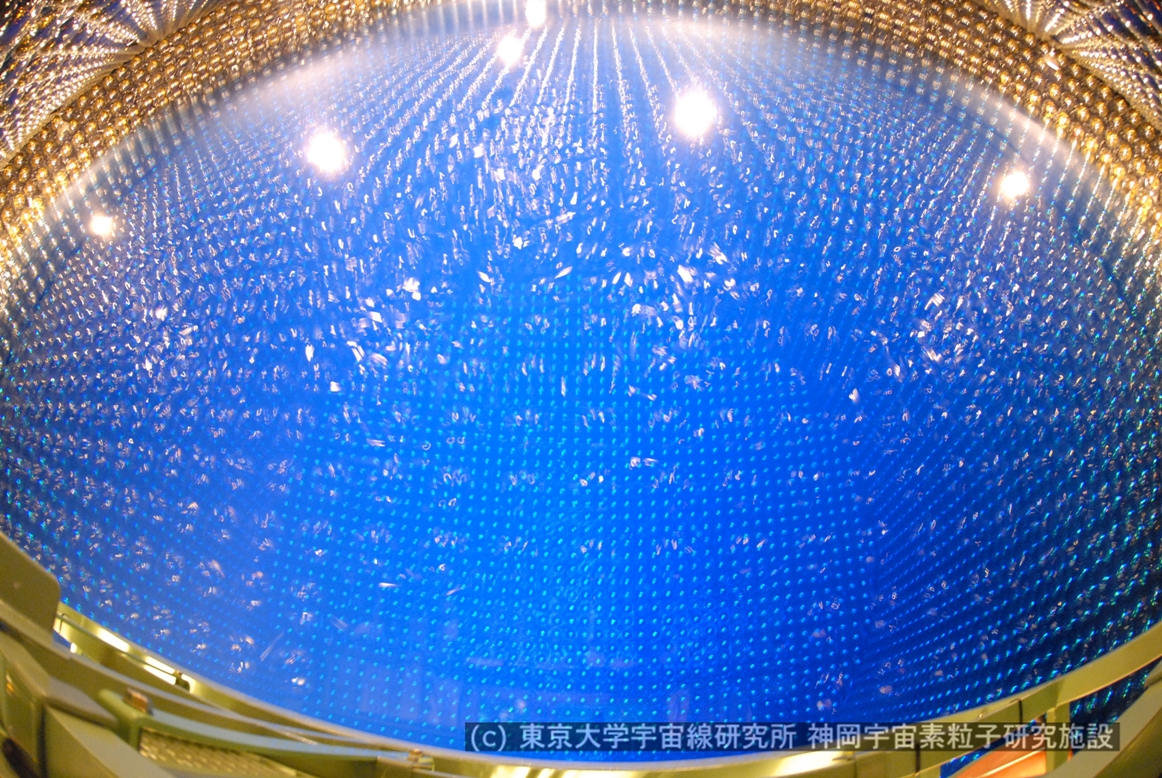} 
  \caption{Super-Kamiokande detector during water filling in June 2006.
  }\label{sk-photo}
 \end{center}
\end{figure}

Super-Kamiokande (SK) \cite{sk-detector} is a cylindrical water Cherenkov
detector with 50-kton of purified water and about 11000 of 20-inch 
PMT's. It is located 1000 m underground (2700 m water equivalence) in
Kamioka Observatory, ICRR, University of Tokyo in Mozumi mine in Gifu
prefecture in Japan.
Figure \ref{sk-photo} shows the inside of the SK detector.
The fiducial volume of the SK detector is 22.5 kton.
The main observation targets of SK detector are solar neutrinos, 
atmospheric neutrinos, supernova neutrinos, and nucleon decays. 
SK is the largest detector in the world for these physics targets. 

SK has started its observation in 1996.
Table \ref{run-hist} shows a summary of the SK experiments.
\begin{table*}
\begin{center}
\begin{tabular}{l l r r c} \hline\hline
Phase  & Run period & Number of ID PMT & Photo coverage & Analysis energy threshold \\ 
       &            &           & ($\%$)         & Total / Kinetic Energy \\ \hline
SK-I   & Apr. 1996 -- Jul. 2001 & 11146  & 40     & 5.0 MeV / 4.5 MeV \\ 
SK-II  & Dec. 2002 -- Oct. 2005 &  5182  & 19     & 7.0 MeV / 6.5 MeV \\ 
SK-III & Jul. 2006 -- Aug. 2008 & 11129  & 40     & 5.0 MeV / 4.5 MeV \\ 
SK-IV  & Sep. 2008 -- today     & 11129  & 40     & (4.5 MeV / 4.0 MeV) \\ 
\hline\hline
\end{tabular}
\end{center}
\label{run-hist}
\caption{Summary of the experimental phases in SK.}
\end{table*}
Currently, SK-IV is running with the lowest energy threshold at the
electron total energy, $E_{total} = 4.5$~MeV. 
In SK-IV, we have replaced the entire front-end electronics system.
All the hit information of PMT's can be read by the electronics
system \cite{qbee}. In SK-IV, we can apply appropriate event time windows, for example,
1.5~$\mu$s for the low-energy (= high rate) events, 
40~$\mu$s for the normal events, 
540~$\mu$s for the high-energy without OD activity events,
and $\pm512$~$\mu$s around the beam spill timing of the T2K experiment.
In future, we would like to lower the energy threshold down to 
$E_{total} = 4.0$~MeV or less. 

\section{Atmospheric neutrino results}

Cosmic-ray interactions in the atmosphere produce neutrinos. 
They are called atmospheric neutrinos.
In the SK detector, atmospheric neutrino events are categorized 
into fully contained (FC) events, partially contained (PC) events, and 
upward going muons.

FC events deposit all of their Cherenkov light in the inner
detector (ID) in SK, while PC events include punch-through tracks which
deposit some Cherenkov light in the outer detector (OD). 
The neutrino interaction vertex is required to be reconstructed within 
the 22.5 kton fiducial volume.
The FC events are classified into ``sub-GeV'' 
(visible energy, $E_{vis} <1330$ MeV) 
and ``multi-GeV'' ($E_{vis} >1330$ MeV). 
These events are further separated into sub-samples based on the number
of observed Cherenkov rings. Single- and multi-ring are then divided into
electron-like (e-like) or muon-like ($\mu$-like) samples depending on pattern
identification of the most energetic Cherenkov ring. 
The sub-GeV samples are additionally divided based on their number of
decay-electrons and their likelihood of being a $\pi^0$.

The PC events are separated into "OD stopping" and "OD through-going"
categories based on the amount of light deposit by the exiting particle
in the OD.

Energetic atmospheric muon neutrinos passing through the Earth interact with
rock surrounding the detector, and produce muons via charged current
interactions. These neutrino events are observed as upward going muons.
Upward going muons are classified into two types. One is
``upward through-going muons'' which have passed through
the detector, and the other is ``upward stopping
muons'' which come into and stop inside the detector.
The upward through-going muons are subdivided into "showering" and
"non-showering" based on whether their Cherenkov pattern is
consistent with light emitted from an electromagnetic shower produced
by a very high energy muon.

The livetime and number of observed events for the first three SK phases
are summarized in \tablename~\ref{tab:fcpcsummary}
and \ref{tab:upmusummary}, respectively.
\begin{table}
\center
\caption{Atmospheric neutrino livetime and the number of observed
FC and PC events for each SK phase.}
\label{tab:fcpcsummary}
\begin{tabular}{crrr}
\hline
 & Livetime (days) & FC & PC \\ \hline
SK-I & 1,489 & 12,232 & 896 \\
SK-II & 799 & 6,584 & 429 \\
SK-III & 518 & 4,356 & 343 \\
\hline
\end{tabular}
\end{table}
\begin{table}
\center
\caption{Atmospheric neutrino induced upward-going muon livetime
and the number of observed events for each SK phase.}
\label{tab:upmusummary}
\begin{tabular}{crrr}
\hline
 & Livetime (days) & through-going & stopping \\ \hline
SK-I & 1,646 & 1,856 & 458 \\
SK-II & 828 & 889 & 228 \\
SK-III & 636 & 735 & 210 \\ \hline
\end{tabular}
\end{table}

\begin{figure}
\begin{center}
\includegraphics[width=8cm]{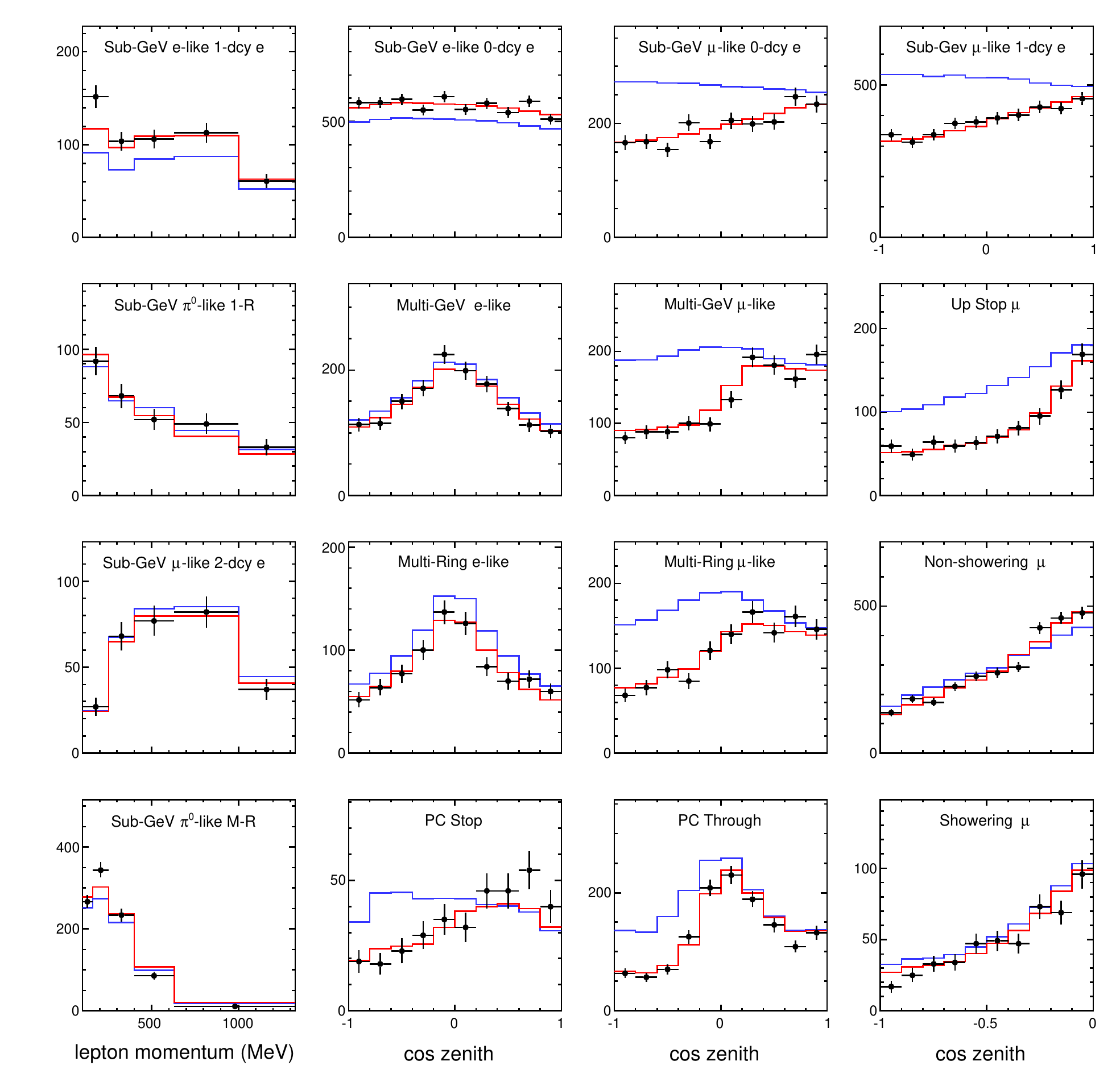}
\caption{The zenith angle and lepton momentum distributions 
for each data sample in SK-I+II+III (preliminary).
$\cos\Theta =$1 indicates downward-going particles.
The blue histograms show the MC prediction without neutrino oscillation
and the red histograms show the MC prediction for $\nu_\mu 
\leftrightarrow \nu_\tau$ oscillations with $\sin^2 2\theta =1.0$ and
$\Delta m^2 = 2.1\times 10^{-3}$~eV$^2$.}
\label{fig:angdist}
\end{center}
\end{figure}

The zenith angle and lepton momentum distributions for each of the above
samples compared with the atmospheric neutrino Monte Carlo (MC)
predictions are shown in Figure~\ref{fig:angdist}. 

\begin{figure} 
\begin{center}
\includegraphics[width=8cm]{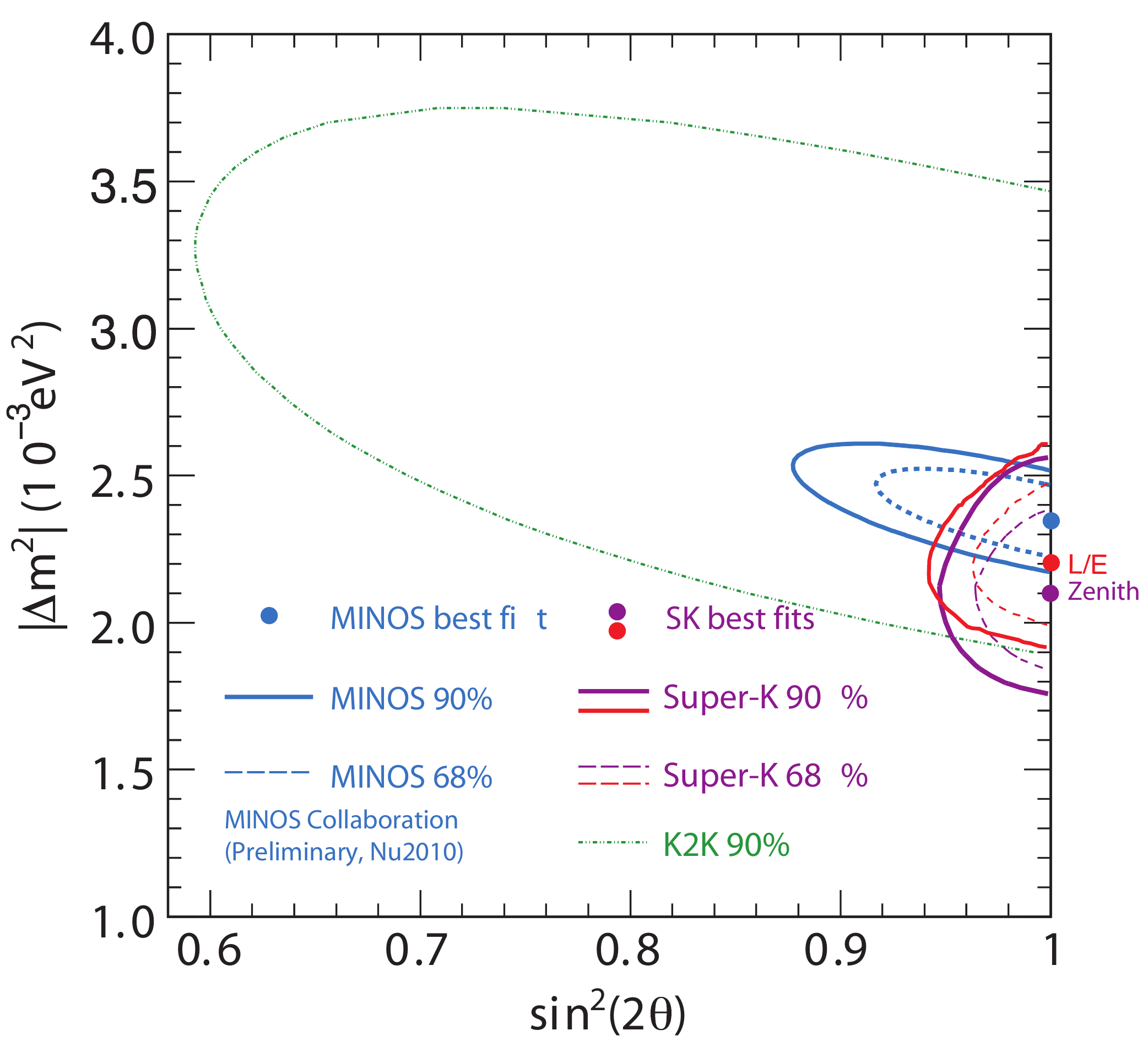}
\end{center}
\caption{Allowed region of $\nu_\mu \rightarrow \nu_\tau$ neutrino 
oscillation parameters obtained by SK using
contained atmospheric neutrino events and upward-going
muon events in SK-I+II+III (preliminary).
Solid and dashed contours correspond to 90 and 68~\% C.L. respectively.
Purple contours are obtained by the zenith angle analysis and
red contours are obtained by the L/E analysis.
Allowed regions by long-baseline neutrino oscillation experiment
K2K and MINOS are shown in green and blue contours, respectively.}
\label{fig:combined}
\end{figure}

In June 2009, we have summarized a preliminary results of 
the neutrino oscillation analysis using SK-I, II and III atmospheric 
neutrino data set.
Red contours in \figurename~\ref{fig:combined} show the allowed neutrino
oscillation parameter regions for $\nu_\mu \leftrightarrow \nu_\tau$ 
oscillations.
The allowed oscillation parameter region from the zenith angle analysis
is $\sin^2 2\theta_{23} > 0.96$ (90~\% C.L.) and 
$\Delta m_{23}^2 =(2.11+0.11-0.19)\times 10^{-3}$~eV$^2$.
The allowed oscillation parameter region from the L/E analysis
is $\sin^2 2\theta_{23} > 0.96$ (90~\% C.L.) and 
$\Delta m_{23}^2 =(2.19+0.14-0.13)\times 10^{-3}$~eV$^2$.

\begin{figure}
\begin{center}
\includegraphics[width=8cm]{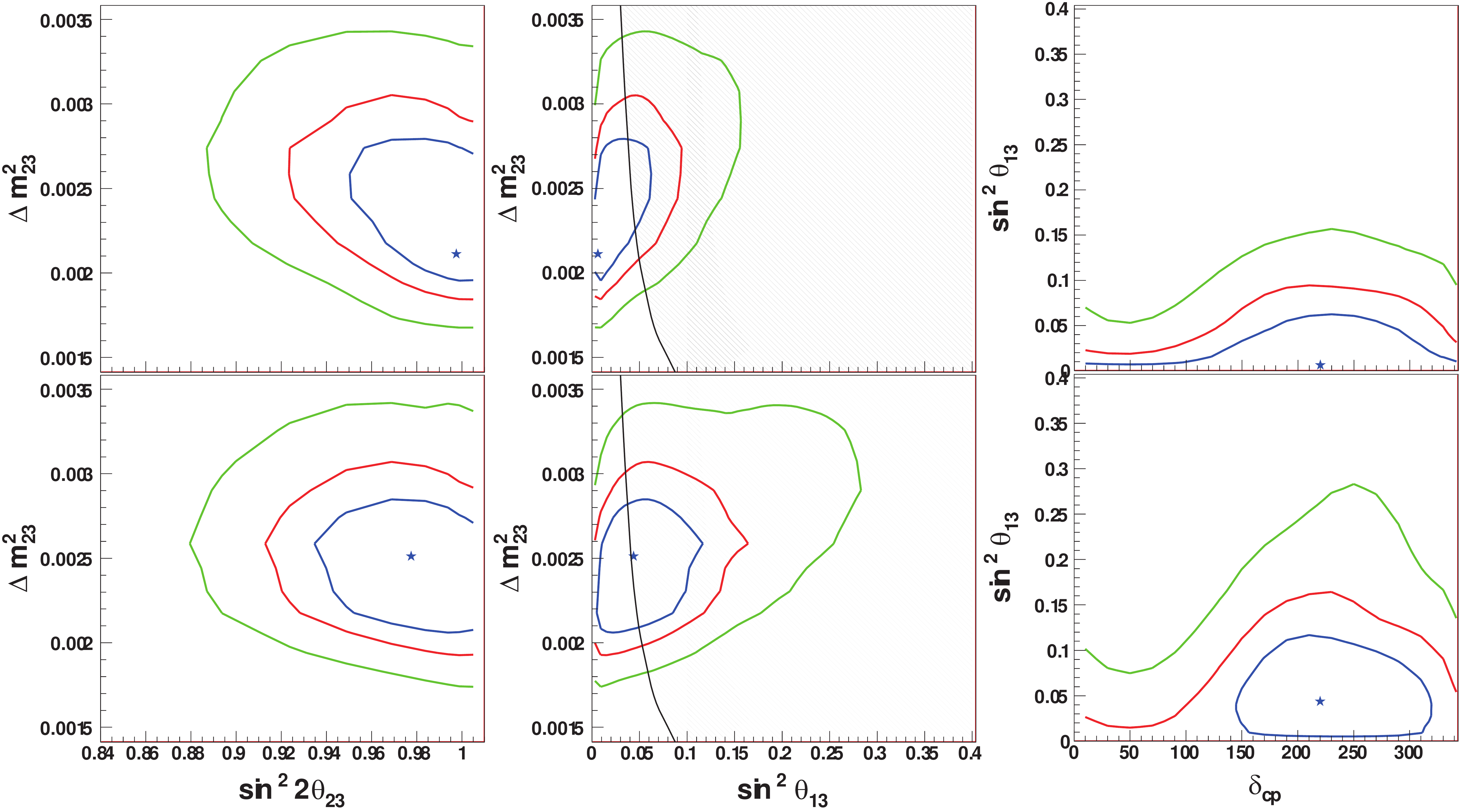}
\caption{The allowed regions for 
($\Delta m_{23}^2$, $\sin^22\theta_{23}$); left,
($\Delta m_{23}^2$, $\sin^2\theta_{13}$); middle, and
($\sin^2\theta_{13}$, $\delta_{cp}$); right
  for the normal (upper figure) and inverted (lower figure) hierarchy
 in SK-I+II+III (preliminary).
The blue, red, and green contours correspond to 68, 90 and 99~\% C.L.
allowed regions obtained by this analysis, respectively.
The shaded regions corresponds
to the area excluded at 90~\% C.L. by the CHOOZ experiment.}
\label{fig:osc3d}
\end{center}
\end{figure}

In May 2010, we have performed a full 3-flavor oscillation analysis
considering all the mixing parameters and the CP violating term.
The matter effect in the Earth is also considered in this calculation
and both the normal and inverted mass hierarchies are tested.
\figurename~\ref{fig:osc3d} shows the allowed regions for 
($\Delta m_{23}^2$, $\sin^22\theta_{23}$),
($\Delta m_{23}^2$, $\sin^2\theta_{13}$), and
($\sin^2\theta_{13}$, $\delta_{cp}$)
for the normal and inverted mass hierarchies with SK-I+II+III data.
A numerical results are summarized in Table \ref{full-3f}.
\begin{table*}
\begin{center}
Normal hierarchy ($\chi^2_{min}=469.94/416$~dof)
\begin{tabular}{c c c c} \hline\hline
Parameter  & Best point & 90~\% C.L. allowed & 68~\% C.L. allowed \\ \hline
$\Delta m_{23}^2(\times 10^3)$  & 2.11 eV$^2$ & 1.88--2.75 eV$^2$ & 1.99--2.54 eV$^2$ \\ 
$\sin^2\theta_{23}$             & 0.525       & 0.406--0.629      & 0.441--0.597 \\ 
$\sin^2\theta_{13}$             & 0.006       & $<$0.066            & $<$0.036   \\
CP-$\delta$                     & 220 $^{\rm o}$ & ---	 & 140.8 $^{\rm o}$--297.3 $^{\rm o}$ \\
\hline\hline
\end{tabular}
\vspace{2mm}
Inverted hierarchy ($\chi^2_{min}=468.34/416$~dof)
\begin{tabular}{c c c c} \hline\hline
Parameter  & Best point & 90~\% C.L. allowed & 68~\% C.L. allowed \\ \hline
$\Delta m_{23}^2(\times 10^3)$  & 2.51 eV$^2$ & 1.98--2.81 eV$^2$ & 2.09--2.64 eV$^2$ \\ 
$\sin^2\theta_{23}$             & 0.575       & 0.426--0.644      & 0.501--0.623 \\ 
$\sin^2\theta_{13}$             & 0.044       & $<$0.122            & 0.0122--0.0850  \\
CP-$\delta$                     & 220 $^{\rm o}$ & 121.4 $^{\rm o}$--319.1 $^{\rm o}$ & 165.6 $^{\rm o}$--280.4 $^{\rm o}$ \\
\hline\hline
\end{tabular}
\end{center}
\label{full-3f}
\caption{Summary of the full 3-flavor oscillation analysis results in
 SK-I+II+III (preliminary). The $\sin^2\theta_{12}$ and $\Delta m_{12}^2$
are fixed at 0.304 and $7.66 \times 10^{-5}$ eV$^2$, respectively.}
\end{table*}
All fits are consistent with the two flavor oscillation results
and CHOOZ experiment's upper limit on $\theta_{13}$.
No preference for either mass hierarchy exists in the data.

We also performed a test for CPT violations using the atmospheric
neutrino data in SK-I+II+III.
We applied the two-neutrino disappearance models for neutrinos
and anti neutrinos, separately:

$P(\nu_\mu\rightarrow\nu_\mu) = 1-\sin^22\theta\sin^2( 1.27 \Delta m^2 L / E )$,

$P(\bar{\nu}_\mu\rightarrow\bar{\nu}_\mu) = 1-\sin^22\bar{\theta}\sin^2(
1.27 \Delta \bar{m}^2 L / E )$.

\begin{figure}
\begin{center}
\includegraphics[width=7.0cm]{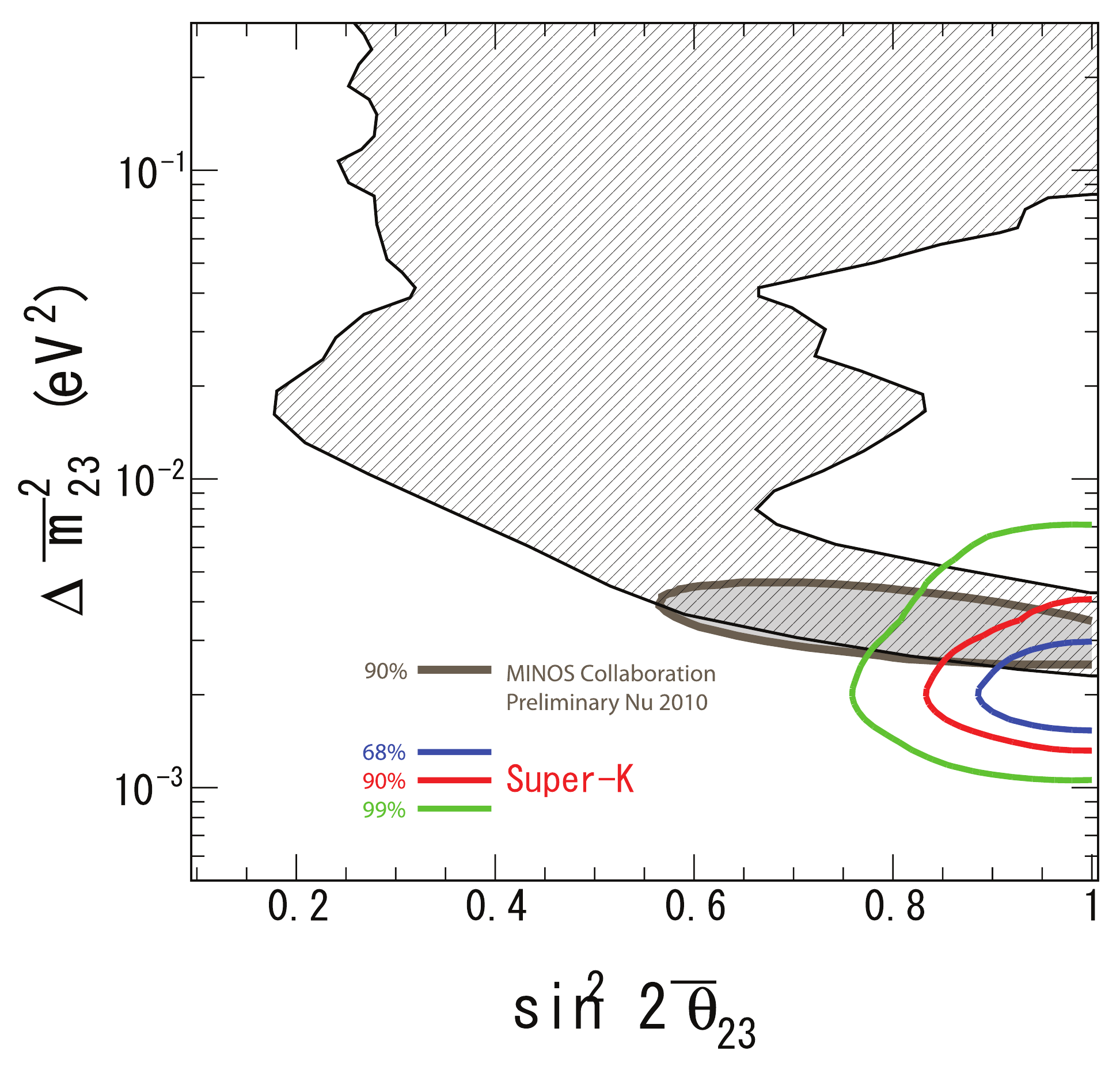}
\caption{Allowed regions for the anti-neutrino mixing parameters in
 SK-I+II+III (preliminary).
The blue, red and green contours correspond to the 68~\%, 90~\%, and 99~\% C.L.
allowed regions, respectively. Shaded region shows the allowed regions by
MINOS experiment.}
\label{fig:cpttest}
\end{center}
\end{figure}

The allowed region for anti-neutrino mixing parameters is shown in 
\figurename~\ref{fig:cpttest}.
The preliminary best-fit parameters are
( $\Delta m^2$, $\Delta \bar{m}^2$, $\sin^22\theta$, $\sin^22\bar{\theta}$ ) =
( $2.2\times 10^{-3}$~eV$^2$, $2.0\times 10^{-3}$~eV$^2$, 1.0, 1.0 ).
The atmospheric mixing parameters for anti-neutrino oscillations are consistent
with those for neutrinos and therefore no evidence for CPT violation is found.

\section{Solar neutrino results}

SK detects solar neutrinos through neutrino-electron elastic scattering (ES), 
$\nu + e \rightarrow \nu + e$, where the energy, direction and time of 
the recoil electron are measured.
Due to its large fiducial mass of 22.5 kton, SK gives a precise
measurement of the solar neutrino flux, energy spectrum, and time
variation via the ES reaction.

In May 2010, we have summarized a preliminary solar neutrino analysis
results in SK-III.
In SK-III, we have re-estimated all the systematic uncertainties.
The preliminary systematic uncertainties on total flux in SK-III are 
summarized in Table \ref{tab:totalsys}.
\begin{table}
\begin{center}
\begin{tabular}{l c  } \hline\hline
Source            &  Total Flux       \\ \hline
Energy scale      &$\pm 1.4   $ \\ 
Energy resolution &$\pm 0.2   $  \\ 
$^8$B spectrum      &$\pm 0.2   $            \\ 
Trigger efficiency        &$\pm 0.5  $   \\ 
Fiducial volume (vertex shift) &$\pm 0.54  $ \\ 
Event Reduction  &$\pm 0.65   $  \\
Small cluster hits cut   &$\pm 0.5   $  \\ 
Spallation        &$\pm 0.2   $ \\ 
External event cut     &$\pm 0.25  $ \\
Background shape  &$\pm 0.1   $  \\ 
Angular resolution   &$\pm 0.67   $ \\ 
Signal extraction method&$\pm 0.7  $   \\ 
Cross section     &$\pm 0.5  $   \\ 
Live time          &$\pm 0.1 $ \\ \hline
Total             &$\pm 2.1  $\\ 
\hline\hline
\end{tabular}
\end{center}
\caption{ Summary of the preliminary systematic uncertainties 
on total flux in $\%$ in SK-III. \label{tab:totalsys}}
\end{table}
The systematic uncertainty on the total flux is estimated to be 2.1 $\%$.
This is about two thirds of the corresponding SK-I value. 
The main contributions to the improvement are
the vertex shift, angular resolution, and event selection uncertainties.

The total live time for solar neutrino analysis in SK-III is 
548 days for $E_{total} \geq 6.5$ MeV, and 289 days for $E_{total} < 6.5$ MeV.
The recoil electron's total energy range used for this analysis 
is from 5.0 to 20 MeV. 
The observed $^8$B solar neutrino flux via the ES reaction is 
$2.32 \pm 0.04 {\rm (stat.)} \pm 0.05 {\rm (syst.)}$ ($\times 10^6/{\rm cm}^2/\rm s$),
which is consistent with the previous SK-I and II results. 
The obtained day-night asymmetry is
$\frac{ (\Phi_{Day} - \Phi_{Night}) }{ (\Phi_{Day} + \Phi_{Night})/2} = -0.056 \pm 0.031 {\rm (stat.)} \pm 0.013 \rm{(syst.)}$.
The total $^8$B and hep flux values are referred from the BP2004 Standard Solar
Model (SSM)~\cite{ssm}. We used the expected spectrum of $^8$B and hep neutrinos 
calculated by Winter~\cite{win06} and Bahcall~\cite{hep}, respectively.

\begin{figure}
\begin{center}
\includegraphics[width=7cm]{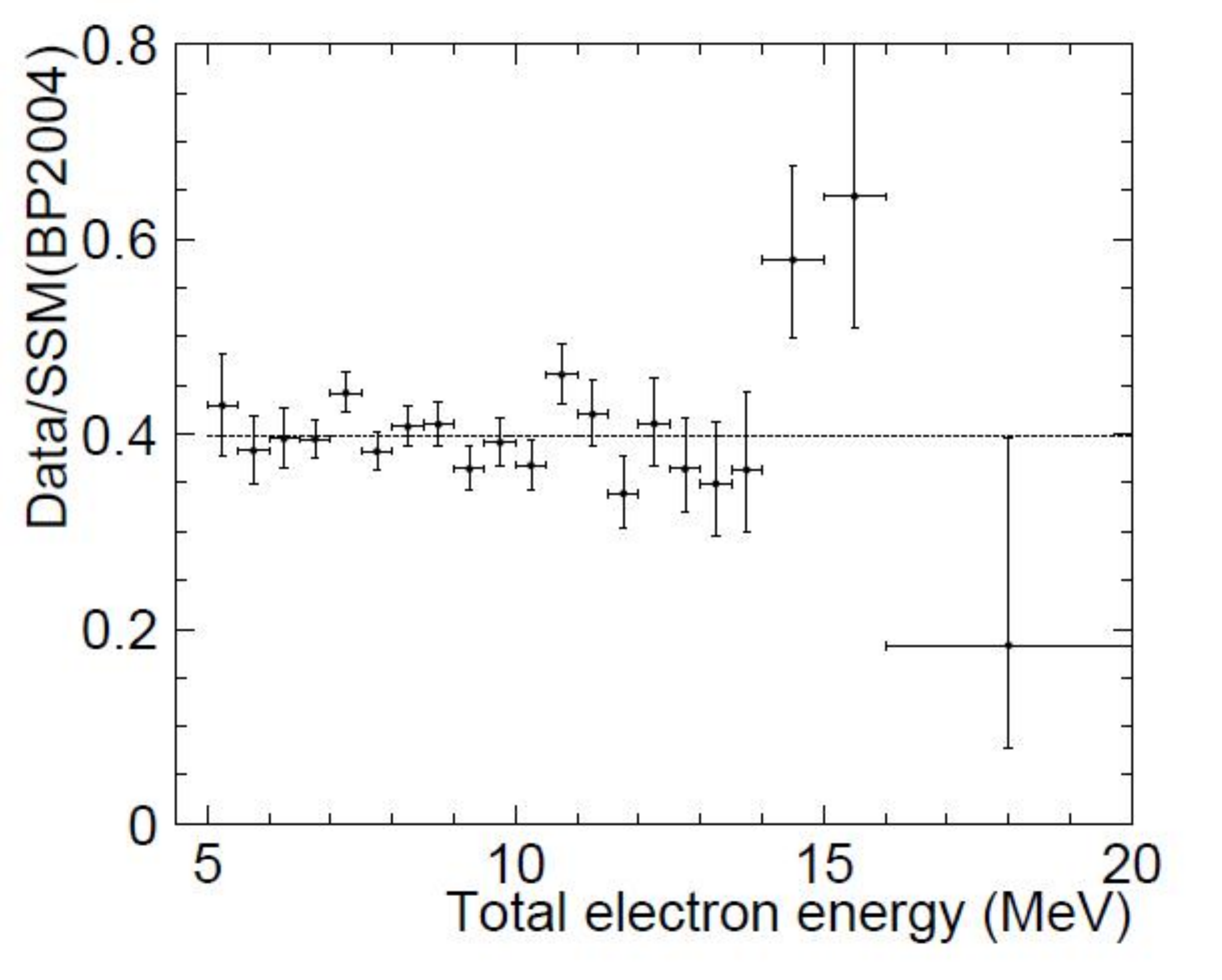}
\caption{Energy spectrum of solar neutrino flux in SK-III data from 5.0 MeV to 20
 MeV (preliminary). Each point shows the ratio of the data and the expected event rate
 calculated from the SSM. 
 The line indicates the averaged value of the SK-III data.}
 \label{fig:sk3energy}
\end{center}
\end{figure}

Figure \ref{fig:sk3energy} shows the observed recoil electron's energy
spectrum divided by the expected signal rate from BP2004 SSM without oscillation.
The line in Figure~\ref{fig:sk3energy} represents the total
SK-III average (flat data/SSM prediction without MSW effect).
This result indicates no significant spectral distortion.

\begin{figure}
\begin{center}
\includegraphics[width=8cm]{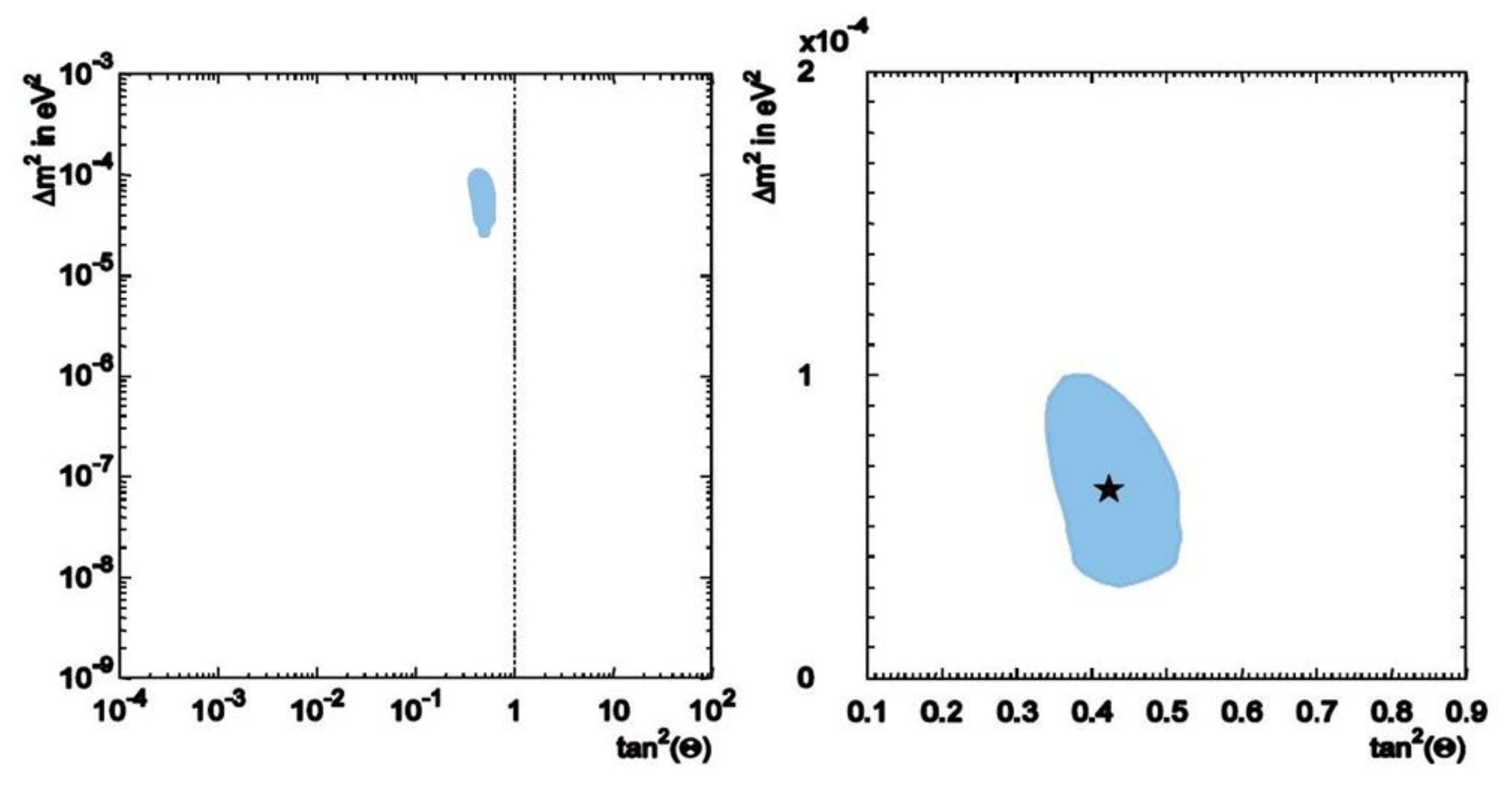}
\caption{ 
Allowed region (95~\% C.L.) for neutrino oscillation parameters, $\Delta m^{2}$ and 
tan$^{2}\theta$ from solar analysis (preliminary). 
The left figure shows the results from 
the SK-I, II and III combined analysis with the $^{8}$B flux constrained
by SNO NC measurements. The right figure presents the result of global 
analysis including SK and other solar experiments 
(SNO, Borexino, Homestake, GALLEX-GNO and SAGE).
The star mark indicates the best-fit parameter set.}
\label{fig:2flavor}
\end{center}
\end{figure}

A preliminary oscillation analysis has also been done 
including the SK-III solar neutrino data. 
The two flavor analysis for the determination of the solar 
oscillation parameters, $\theta_{12}$ and $\Delta m^{2}_{12}$ was 
performed with the combined SK-I, II and III data. 
In this analysis, the total $^{8}$B flux is constrained by 
the averaged NC flux from SNO NCD~\cite{snoncdn} and LETA~\cite{snoleta} 
($ = (5.14 \pm 0.21) \times 10^6/{\rm cm}^2/\rm s$).
The result is shown in the left side in Figure \ref{fig:2flavor}. 
As shown in the left figure, the LOW solution is excluded and the LMA
solution is only allowed to explain the solar neutrino oscillation. 
We have also done a preliminary global oscillation analysis with other solar
experiments.
In this analysis, the following observation results are used; 
SNO total CC rates observed in the 306-day pure D$_2$O phase
(SNO-I)~\cite{snod2o},
391-day salt phases~(SNO-II)~\cite{snosalt}, 
and 385-day NCD phase (SNO-III)~\cite{snoncdn},
the combined NC rates of LETA~\cite{snoleta} and SNO-III,
the predicted day-night asymmetry for SNO-I and II,
$^7\mathrm{Be}$ solar neutrino flux from Borexino~\cite{borexino},
Homestake~\cite{cl},
GALLEX-GNO~\cite{gallex_gno},
 and SAGE~\cite{sage}.
The result is shown in the right side in Figure \ref{fig:2flavor}. 
The best-fit parameter set is $\tan^2\theta_{12} = 0.42^{+0.04}_{-0.02}$
and 
$\Delta m^2_{21} = (6.2^{+1.1}_{-1.9})\times 10^{-5}$~eV$^2$.
In addition, combining the above and KamLAND data~\cite{kamland}, 
the best-fit parameter set becomes 
$\tan^2\theta_{12} = 0.44 \pm 0.03$ and 
$\Delta m^2_{21} = (7.6 \pm 0.2) \times 10^{-5}$~eV$^2$.

We also performed a preliminary 3-flavor oscillation analysis.
In the 3-flavor analysis, the calculation of oscillation probability is
based on reference~\cite{barger}.
The probability can be calculated with three parameters:
$\theta_{12}$, $\theta_{13}$, and $\Delta m^2_{12}$, 
assuming $\Delta m^2_{12} \ll \Delta m^2_{23} \sim \Delta m^2_{13}$. 
We fixed $\Delta m^2_{23} = 2.4 \times 10^{-3}$~eV$^2$ and 
the normal hierarchy is assumed. 
For the solar neutrino oscillation, the other mixing parameters 
are irrelevant, but we set $\theta_{23}=\pi/4$ and $\delta_{CP}=0$ 
in our calculation.

As done for the two-flavor analysis, the oscillation probabilities
depending on different zenith angles are calculated. 
Other experiment data are have been published by Homestake, SAGE,
Gallex-GNO, Borexino, SNO, and KamLAND.
The oscillation parameters are scanned in the following regions:
$10^{-5}$~eV$^2 <\Delta m^2_{12} < 2 \times 10^{-4}$~eV$^2$, $0.1< \tan^2
\theta_{12}< 1.0 $, and $0< \sin^2\theta_{13}< 0.25$. 

\begin{figure}
\begin{center}
\includegraphics[width=6cm]{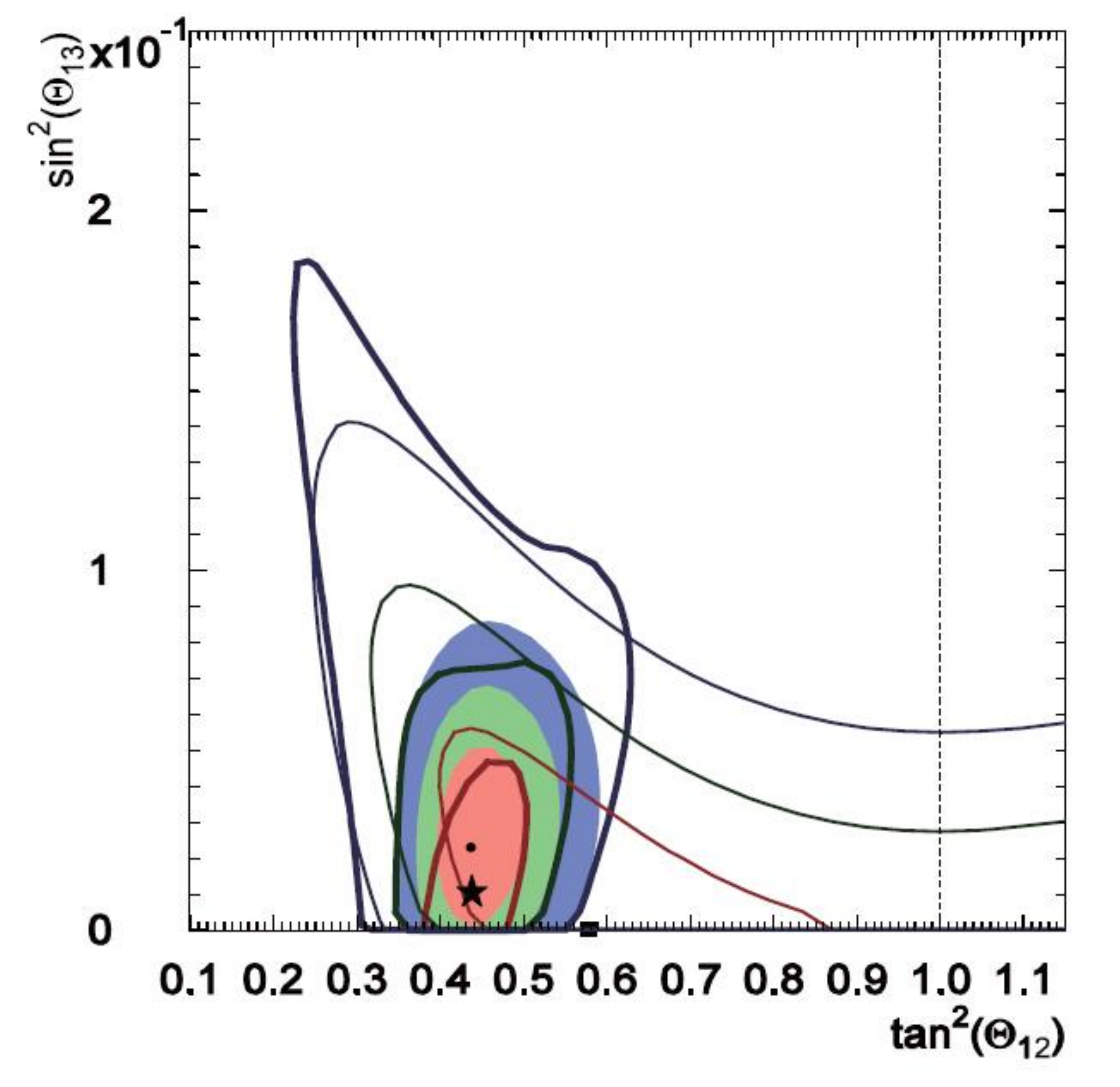}
\caption{
Allowed region for sin$^{2}\theta_{13}$ and tan$^{2}\theta_{12}$ from
 the preliminary 3-flavor analysis.
The thick lines and the star mark show the allowed regions and 
the best-fit point of the solar global analysis.
The thin lines and the square mark show the allowed regions and 
the best-fit point of our KamLAND analysis. 
The filled areas and the filled circle mark show the allowed 
regions and the best-fit point of the combined analysis. 
For all regions, the innermost area (red), the middle area (green) 
and the outermost area (blue) show 68.3, 95, 99.7~\% C.L. respectively.
}
\label{fig:3flavor}
\end{center}
\end{figure}

Figure \ref{fig:3flavor} shows the allowed region for
sin$^{2}\theta_{13}$ and tan$^{2}\theta_{12}$ extracted from this analysis. 
The allowed region from the combined analysis of 
solar global and KamLAND analyses is also indicated. 
The global solar analysis finds that the upper limit 
of sin$^{2}\theta_{13}$ is 0.060 at 95~\% C.L.
After combining with the KamLAND result, the best fit value
 of sin$^{2}\theta_{13}$ is found to be $0.025^{+0.018}_{-0.016}$ 
and the upper limit is obtained as sin$^{2}\theta_{13} < 0.059$ at 
95~\% C.L. 

\section{Conclusion}

Super-Kamiokande-IV is running with the lowest energy threshold in SK.
Currently, 100~\% efficiency is at $E_{total}=4.5$ MeV.

Preliminary results of the full 3-flavor atmospheric neutrino
oscillation analysis from SK-I+II+III data are obtained.
There is no preference for either mass hierarchy in the data, 
and no significant constraint applied on CP phase at 90~\% C.L.

A CPT violation study on atmospheric neutrino is also done with 
SK-I+II+III data. No evidence for CPT violation is found, but
this is preliminary.

Preliminary results of the solar neutrino measurement in SK-III are
obtained. The measured solar $^8$B flux via ES reaction is 
$(2.32\pm0.04(\textrm{stat.})\pm
0.05(\textrm{sys.}))\times10^6~\textrm{cm}^{-2}\textrm{s}^{-1}$.
In a preliminary 3-flavor global solar neutrino oscillation analysis
with KamLAND,
the best-fit value of $\sin^2 \theta_{13}$ is found to 
be $0.025^{+0.018}_{-0.016}$ and an upper bound is obtained 
as $\sin^2\theta_{13} < 0.059$ at 95~\% C.L.  

\vspace{5mm}
{\bf Acknowledgments}
\vspace{5mm}

The author gratefully acknowledges the cooperation of 
the Kamioka Mining and Smelting Company. 
Super-Kamiokande has been built and operated from funds provided 
by the Japanese Ministry of Education, Culture, Sports, Science and
Technology, 
the U.S. Department of Energy, and the U.S. National Science
Foundation. 
This work was partially supported by the  Research Foundation of 
Korea (BK21 and KNRC), the Korean Ministry of Science and Technology, 
the National Science Foundation of China, and 
the Spanish Ministry of Science and Innovation 
(Grants FPA2009-13697-C04-02 and Consolider-Ingenio-2010/CPAN).



\vspace{5mm}
{\bf References}
\vspace{5mm}






\end{document}